\documentclass[sigconf,screen]{acmart}
\AtBeginDocument{
  }
\usepackage{booktabs}
\usepackage{tabularx}
\usepackage{enumitem}
\usepackage{listings}
\providecommand{\correspondingauthor}{\authornote{Corresponding author.}}
\lstdefinelanguage{JavaScript}{
  keywords={async, await, try, catch, function, return, const, let},
  sensitive=true, comment=[l]{//}, string=[b]', string=[b]"
}
\setcopyright{none}
\acmDOI{10.1145/3843779.3844632}
\acmYear{2026}
\copyrightyear{2026}
\acmISBN{979-8-4007-2991-1/2026/10}
\acmConference[POVC '26]{Proceedings of the 1st International Workshop on PromptOps and Vibe Coding}{October 12--16, 2026}{Munich, Germany}
\acmBooktitle{Proceedings of the 1st International Workshop on PromptOps and Vibe Coding (POVC '26), October 12--16, 2026, Munich, Germany}
\acmSubmissionID{asews26povcmain-p7-p}
\received{2026-08-05}
\received[accepted]{2026-08-23}

\newcounter{prompt}
\newenvironment{promptbox}[1]{%
  \refstepcounter{prompt}
  \begin{center}
  \begin{minipage}{0.96\linewidth}
  \hrule
  \smallskip
  \noindent\textbf{#1}
  \par\smallskip
}{%
  \par\smallskip
  \hrule
  \end{minipage}
  \end{center}
}
\title[VibeCheck: Assessing the Quality of LLM-Generated Unit Tests]{VibeCheck: Assessing the Quality of LLM-Generated Unit Tests: A Multi-agent Empirical Study across Heterogeneous Repositories}

\author{Anika Tabassum}
\orcid{0009-0008-3697-0304}
\affiliation{%
  \institution{University of Dhaka}
  \city{Dhaka}
  \country{Bangladesh}
}
\email{anika-2019417844@cs.du.ac.bd}

\author{Mushahid Intesum}
\orcid{0009-0005-7764-9355}
\affiliation{%
  \institution{University of Dhaka}
  \city{Dhaka}
  \country{Bangladesh}
}
\email{intesummushahid@gmail.com}

\author{Md. Fahim Arefin}
\correspondingauthor
\orcid{0000-0002-3413-4774}
\affiliation{%
  \institution{University of Dhaka}
  \city{Dhaka}
  \country{Bangladesh}
}
\email{fahim@cse.du.ac.bd}

\author{Tarannum Shaila Zaman}
\orcid{0000-0002-8634-524X}
\affiliation{%
  \institution{University of Maryland}
  \city{Baltimore County}
  \country{USA}
}
\email{zamant@umbc.edu}

\newcommand{\Name}{VibeCheck}
\begin{document}
\begin{abstract}
LLM-based IDE agents are increasingly used to generate repository-grounded unit tests, yet common evaluations often rely on execution success or coverage. These metrics can miss deeper quality issues such as weak assertions, missing edge cases, poor isolation, and limited maintainability. This paper presents \Name{}, an empirical study of unit test generation across 15 student-developed Python and JavaScript/TypeScript repositories. We evaluate Kiro, Antigravity, and Cursor with Claude Sonnet 4.5 as the underlying agent, under repository-only, zero-shot conditions using a five-dimensional rubric covering runnability, assertion strength, logic and edge-case coverage, isolation/determinism, and maintainability. We also apply leave-one-out cross-agent peer evaluation to compare tools and identify failure patterns. Results show a clear execution-adequacy gap: generated tests are often runnable but frequently lack strong assertions and meaningful behavioral coverage. Weak assertions and missing edge cases occur more often than blocking failures, showing that runnable tests can still be shallow. \Name{} provides a reliability-oriented framework for evaluating LLM-generated tests beyond pass/fail outcomes.
\end{abstract}



\maketitle

\section{Introduction}
\label{intro}

Large Language Model (LLM)-based IDE agents are increasingly used to generate unit tests from repository context, yet developers and researchers lack lightweight, reusable support for inspecting whether these tests are merely executable or actually useful. A suite may run successfully while relying on weak assertions, missing edge cases, introducing flaky behavior, or failing to remain maintainable~\cite{rothermel1996regression,fraser2015fifty,10.1145/3803437.3805561}---making execution success alone an incomplete signal of quality~\cite{Zhang2024TestBench}.

Unit testing validates the smallest independently testable components of a program in isolation~\cite{Mundler2024SWT}, with each test case defining an input, executing the target functionality, and checking observed behavior through assertions~\cite{Jain2024TestGenEval}. Well-designed tests improve reliability, support regression testing, and serve as executable documentation~\cite{schafer2023llm-tests},\cite{tang2023evaluation}.

To support more systematic assessment, we present \Name{}, a repository-grounded evaluation artifact and reusable framework for analyzing LLM-generated unit test quality. It provides a controlled workflow for collecting generated tests, applying cross-agent peer evaluation, scoring suites with a five-dimensional rubric, and summarizing diagnostic failure patterns---targeting practical questions for teams integrating AI coding agents into software maintenance: which IDEs generate broader suites relative to a common baseline, which outputs are more runnable, where adequacy breaks down, and whether evaluator confidence can be trusted.

We apply \Name{} to 15 student-developed Python and Java\-Script/TypeScript repositories using three AI-assisted IDEs---Kiro, Antigravity, and Cursor---each generating tests under zero-shot, repository-only constraints, with no external documentation access and no production-code modification. All three use Claude Sonnet 4.5 as their underlying model; as our baseline, we use standalone Claude Sonnet 4.5 (CS-4.5) to generate tests from the same repository-only prompt without an IDE wrapper. We compare against this baseline because each IDE handles repository context differently even on the same underlying model, producing measurably different results that a bare model comparison alone would not reveal. Suites are assessed on runnability, assertion strength, logic and edge-case coverage, isolation/determinism, and maintainability, alongside diagnostic signals such as blocking errors, flakiness risks, weak assertions, and missing behavioral cases.

Our results expose a consistent gap between runnable and behaviorally adequate tests, holding for both the IDEs and the CS-4.5 baseline: executable tests are common, but recurring weaknesses remain in assertion quality, edge-case coverage, isolation, and maintainability. \Name{} also supports cross-agent peer evaluation, enabling comparison of evaluator consistency and confidence-quality alignment relative to the baseline.

Our generation prompt (Prompt 1), evaluation prompt (Prompt 2), repository-only interaction protocol, and scoring rubric are fixed, versioned, and released alongside the results, enabling reproducible comparison of AI coding IDEs against a common baseline. With the emerging prominence of PromptOps\footnote{A set of practices that applies DevOps-style principles to maintain prompts and tools used with LLMs in the development cycle}, we design \Name{} so teams can integrate these prompts directly into their existing DevOps\footnote{Practices that automate and integrate software development, testing, and deployment for faster, more reliable releases} pipelines.

In this paper, we make the following contributions:

\begin{itemize}
\item \Name{} is a reusable, execution-grounded evaluation framework for repository-grounded LLM unit-test generation, combining a five-dimensional adequacy rubric, developer-defined behavioral obligations, and leave-one-out cross-agent peer evaluation.
\item We provide an empirical study of three contemporary AI-assisted IDEs (Kiro, Antigravity, and Cursor), each built on CS-4.5, benchmarked against standalone CS-4.5 across 15 heterogeneous Python and JavaScript/TypeScript repositories, quantifying the gap between runnability and behavioral adequacy and characterizing dominant failure modes.
\item We release a public replication package containing all prompts, generated test suites, rubric scores, peer-evaluation JSON reports, and analysis scripts, enabling teams to integrate these prompts into their existing DevOps pipelines.
\end{itemize}

\section{Motivation and Background}

AI-assisted IDE adoption has grown rapidly in development teams~\cite{jetbrains2026devsurvey,stackoverflow2025devsurvey}, driven by multi-file editing, codebase-aware context retrieval, and background task execution. Antigravity centers on multi-agent orchestration, where multiple agents plan, execute, and verify tasks autonomously, surfacing approval requests and Artifacts (task lists, implementation plans, browser recordings) without requiring review of raw tool calls. Kiro centers on spec-driven governance, translating requirements into structured specifications before code generation, with event-driven hooks and repository-level steering files enforcing testing and coding standards as files change. Cursor centers on deep codebase context via a semantic vector index of the project, exposing graduated autonomy through Tab completions, Composer multi-file edits, and a longer-running Agent Mode that edits files and runs terminal commands iteratively.

Practical deployment exposes an important reliability concern. Consider a developer generating tests for a newly implemented authentication component before opening a pull request. The suite compiles, passes, and achieves high coverage---suggesting the implementation is well-verified. Yet inspection reveals that the tests mostly assert non-null return values, omit invalid-credential and session-expiration scenarios, reuse shared mutable fixtures, and duplicate setup code that complicates maintenance---leaving regressions such as incorrect authorization decisions undetected until deployment despite satisfying execution-based metrics. This gap between execution success and behavioral adequacy, and between IDEs built on the same underlying model, motivates the repository-grounded evaluation framework described next.

\section{Related Work}

Automated unit test generation has progressed from random testing, symbolic execution, and search-based software testing (SBST)---tools such as EvoSuite and Pynguin optimize statement and branch coverage~\cite{fraser2015fifty,bhatia2023pynguin}---toward LLM-based, semantics-aware generation that leverages learned programming knowledge to produce more natural, developer-like tests~\cite{Hossain2025LLM,Tabassum2026ProS}. However, SBST tools often yield brittle tests with weak oracles and poor maintainability~\cite{Pecorelli2022Toward}, and even LLM-based approaches, while achieving competitive or higher coverage than traditional tools~\cite{siddiq2023llm-testgen,schafer2023llm-tests,bhatia2023pynguin}, frequently produce assertions that are incorrect or superficial---showing that execution success alone is insufficient for judging quality. Yuan et al. showed ChatGPT-generated tests are more natural and maintainable than traditional techniques but still require substantial repair before practical use~\cite{Yuan2023No}; ChatUniTest improved branch coverage and assertion quality by combining repository-context extraction with validation-and-repair loops~\cite{Chen2023ChatUniTest}. Broader empirical studies confirm that prompting strategy, model choice, and repository context substantially affect executability and coverage, while behavioral-reasoning weaknesses persist~\cite{yang2024prompt-impact}.

Subsequent work has targeted these weaknesses through richer repository context, prompt engineering, retrieval augmentation, execution feedback, iterative refinement, and multi-agent collaboration~\cite{zheng2023llmjudge}. Context-aware prompting incorporating project structure and dependency information improves executability and coverage~\cite{konuk2024evaluating,yang2024prompt-impact,chang2025benchmark}, while feedback-driven approaches---execution-guided refinement, critique loops, and fine-tuning---further improve compilation success, mutation scores, and overall quality~\cite{jain2025testforge,rehan2025focal-methods}. More recent frameworks incorporate retrieval-augmented generation and richer program analysis to improve assertion generation, type correctness, and behavioral coverage in real-world repositories~\cite{huang2025ult}.

Despite these advances, existing work primarily evaluates generated tests using compilation success, execution rate, coverage, or mutation scores on curated benchmarks and mature repositories~\cite{dinella2022toga,foster2025ach}, with comparatively little attention to repository-grounded IDE agents under strict repository-only constraints that jointly consider assertion strength, behavioral adequacy, isolation, determinism, maintainability, and alignment with developer-expected behavior~\cite{hossain2023oracle}. Few studies examine cross-agent peer evaluation or whether evaluator confidence correlates with observed quality~\cite{luo2014flaky}. Our work addresses these gaps by introducing a repository-grounded evaluation framework combining a five-dimensional adequacy rubric, developer-defined behavioral obligations, and leave-one-out cross-agent peer evaluation across heterogeneous multi-file software projects.

\section{Methodology}
\label{sec:methodology}

\begin{figure*}[t]
  \centering
  \includegraphics[width=\textwidth]{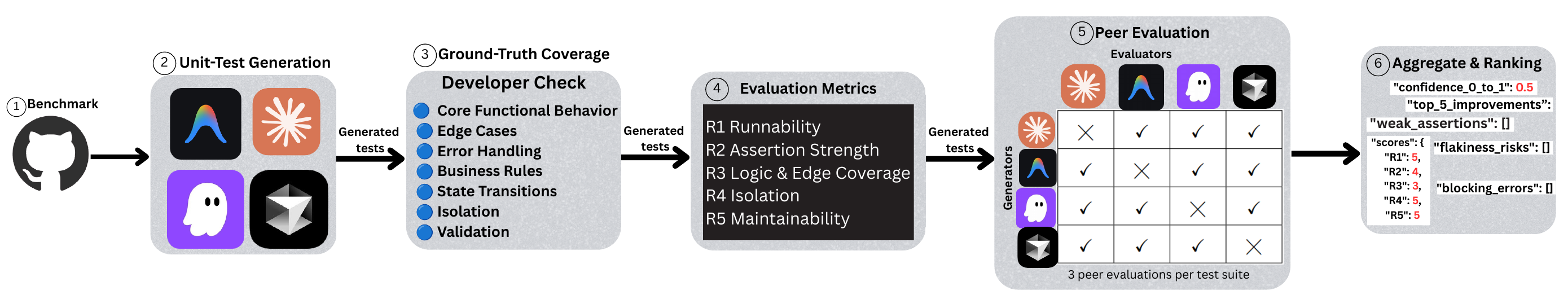}
  \caption{Overview of the \Name{} evaluation pipeline.}
  \Description{A six-stage workflow from benchmark construction and unit-test generation through ground-truth coverage checking, evaluation metrics, cross-agent peer evaluation, and result aggregation.}
  \label{fig:workflow}
\end{figure*}

\Name{} is a controlled framework for evaluating repository-grounded unit test generation by AI-assisted IDE agents. Rather than proposing a new test-generation model, \Name{} provides a systematic methodology for assessing how reliably existing agents generate executable and behaviorally meaningful unit tests under repository-only constraints. In this setting, a generated test suite is considered useful only if it both executes successfully within the target repository and contains assertions that meaningfully validate the intended program behavior, rather than merely increasing the number of test cases.

Figure~\ref{fig:workflow} illustrates the overall evaluation pipeline. \Name{} consists of six stages: benchmark construction, unit-test generation, ground-truth coverage checking, evaluation metrics, cross-agent peer evaluation, and result aggregation. Together, these stages enable a comprehensive assessment of AI-generated unit tests beyond conventional execution-based metrics by examining assertion quality, behavioral coverage, isolation, determinism, and maintainability.

\subsection{Benchmark}

We construct a benchmark of 15 student-developed GitHub repositories spanning the Python and JavaScript/TypeScript ecosystems, covering diverse application domains including backend services, full-stack web applications, single-page applications, and a browser extension. Each repository is cloned and preserved in its original form---source code, dependency manifests, configuration files, and any existing testing infrastructure intact.

Repositories are selected to represent coursework-scale yet non-trivial systems: we include only multi-file projects with identifiable functional components and meaningful business logic, excluding toy examples and single-file assignments. This design evaluates whether AI-assisted IDE agents can reason over realistic repository structures, project-specific implementation details, and framework conventions, rather than solving isolated programming tasks. Although modest in size, the benchmark was curated to span diverse repository structures, languages, and domains, enabling detailed manual assessment while keeping the evaluation workload feasible.

As these repositories are student-authored coursework projects, we cannot publicly release the underlying source code due to privacy and academic-integrity/licensing constraints governing student work; accordingly, our replication package includes the generation and evaluation prompts, generated test suites, rubric scores, peer-evaluation JSON reports, and analysis scripts, but not the original repositories themselves. As a baseline against which each IDE's tests are evaluated, we use the unit tests and evaluation results produced by the standalone CS-4.5 model on the same repositories.

\subsection{Unit-Test Generation}

For each repository, four agents---Kiro, Antigravity, Cursor, and CS-4.5---generate unit tests using the same zero-shot instruction shown in Prompt~\ref{prompt:testgen}. The three IDEs use CS-4.5 as their underlying model, with the standalone CS-4.5 results serving as the baseline. Although the IDEs share the same underlying model, they produce different results because of their differing context-handling strategies. Agents are provided access only to repository-internal files and are prohibited from using external documentation, assuming undocumented behavior, or modifying production code.
\begin{promptbox}{Prompt 1: Unit Test Generation}\label{prompt:testgen}
\footnotesize
\textit{You are given access to the entire project directory.}\par\smallskip

\textbf{Generate unit test cases for this project under the following constraints:}
\begin{itemize}[leftmargin=*, nosep]
  \item Use only the information available in the project files.
  \item Do not assume undocumented behavior.
  \item Do not modify production code.
  \item Follow existing language, framework, and conventions.
  \item Focus on correctness, edge cases, and failure scenarios.
  \item Avoid unnecessary mocking.
\end{itemize}

\textbf{Output:}
\begin{itemize}[leftmargin=*, nosep]
  \item Only unit test source code.
  \item No explanations or additional text.
\end{itemize}

\textbf{Do not request additional input.}
\end{promptbox}

Generation proceeds file-by-file rather than as a single batch pass: for each source file the agent identifies as testable, it generates a corresponding set of test cases, executes them against the target repository, and inspects the resulting output before proceeding to the next file. This per-file generate-run-check cycle allows agents to surface immediate execution failures (e.g., import errors, missing fixtures, dependency issues) within their own workflow before the suite is handed off for evaluation. All generated outputs are preserved exactly as produced at the end of this process, without manual editing, repair, or filtering, ensuring that the evaluation faithfully reflects each agent's repository-grounded test-generation capability.

\subsection{Ground-Truth Coverage}

While peer evaluation measures overall test-suite quality, it does not verify coverage of the repository's expected behaviors. To complement the rubric, we construct a repository-specific checklist of testing obligations for each project, derived independently by the authors from each repository's source code, documentation, and implementation logic---business rules, input/output contracts, and edge cases---\emph{before} and without reference to any generated test suite, ensuring obligations reflect developer-intended behavior rather than the outputs under evaluation.

Obligations cover normal functionality, edge cases, error handling, input validation, state transitions, and project-specific business rules, each mapped to its corresponding module, function, or user-facing behavior. A suite is credited with satisfying an obligation only if at least one test exercises that behavior with a meaningful assertion, preventing superficial tests (e.g., non-null checks) from counting as adequate coverage. This analysis thus complements the peer-evaluation rubric by measuring alignment with developer-expected functionality.

\subsection{Evaluation Metrics}

Each generated test suite is evaluated using a five-dimensional rubric designed to measure both executability and behavioral adequacy. Rather than relying solely on execution success or code coverage, the rubric assesses whether generated tests meaningfully validate program behavior and adhere to good testing practices.

\subsubsection*{\textbf{Runnability}} Measures whether the generated tests are likely to execute successfully without import, dependency, fixture, or configuration errors.
\subsubsection*{\textbf{Assertion Strength}} Evaluates whether assertions verify meaningful program behavior instead of relying on weak, tautological, or superficial checks.
\subsubsection*{\textbf{Logic and Edge-case Coverage}} Assesses whether the generated tests exercise normal execution paths, boundary conditions, failure scenarios, and other important behavioral cases.
\subsubsection*{\textbf{Isolation and Determinism}} Measures whether tests remain independent of external resources such as databases, networks, randomness, or system time, thereby reducing the risk of flaky behavior.
\subsubsection*{\textbf{Maintainability}} Evaluates the readability, organization, modularity, and consistency of the generated tests with the repository's existing testing conventions.

In addition to the rubric scores, evaluators record several diagnostic signals, including blocking errors, flakiness risks, weak assertions, missing behavioral cases, suggested fixes, an overall quality score, and evaluator confidence. These complementary diagnostics provide greater insight into the causes of low-quality test suites, allowing us to distinguish between suites that fail to execute and those that execute successfully but provide limited behavioral validation.

\subsection{Cross-Agent Peer Evaluation}

Each generated test suite is evaluated by the remaining agents using a leave-one-out protocol, ensuring that no agent evaluates its own output. Consequently, every generated test suite receives three independent peer reviews. This evaluation strategy reduces potential self-preference while enabling a more balanced assessment across different LLM-based IDE agents. As shown in Prompt~\ref{prompt:peereval}, evaluators are instructed to act as strict repository-grounded unit-test reviewers and to assess the generated tests solely using the repository context and the generated test files.

\begin{promptbox}{Prompt 2: Cross-Model Peer Evaluation}\label{prompt:peereval}
\scriptsize
\textbf{You are a strict unit-test evaluator for a software testing benchmark.}

\textbf{Goal:}
Given A project context and B AI-generated test files, you will judge test quality using a consistent rubric, \textbf{WITHOUT} assuming missing code works.

\textbf{You MUST:}
\begin{itemize}[leftmargin=*, nosep]
  \item Detect issues that would prevent the tests from running.
  \item Judge whether assertions test real behavior or are superficial.
  \item Judge isolation: DB/network/time/random/external services must be mocked or justified.
  \item Identify flakiness risks.
  \item Identify missing edge cases for the functions/classes being tested.
  \item Provide actionable fixes and improved test examples when possible.
\end{itemize}

\textbf{Rubric (0-5 each):}
\begin{itemize}[leftmargin=*, nosep]
  \item R1 Runnability
  \item R2 Assertion strength
  \item R3 Coverage of logic/edges
  \item R4 Isolation \& determinism
  \item R5 Maintainability/readability
\end{itemize}

\textbf{Output:} JSON report containing file-level scores, blocking errors, flakiness risks, weak assertions, missing cases, suggested fixes, overall score, and evaluator confidence.

\textbf{Inputs:}
\begin{itemize}[leftmargin=*, nosep]
  \item \texttt{PROJECT\_CONTEXT}: key files/snippets from the repository.
  \item \texttt{TEST\_FILES}: generated unit test file contents.
\end{itemize}
\end{promptbox}

Each evaluator assigns scores using the five-dimensional evaluation rubric described in Section~\ref{sec:methodology} and produces a structured JSON report containing rubric scores, diagnostic findings, an overall quality assessment, and evaluator confidence. The resulting peer evaluations provide complementary perspectives on the strengths and weaknesses of each generated test suite while maintaining a consistent evaluation protocol across all repositories and agents.

\subsection{Aggregation and Robustness Analysis}

For each repository-generator pair, we aggregate the evaluations from the three peer reviewers. Let $\mathbf{s}_{p,g,e} \in [0,5]^5$ denote the five-dimensional rubric score assigned by evaluator $e$ to the test suite generated by agent $g$ for repository $p$. The peer-aggregated rubric is calculated using Equation~\ref{eq:peer-aagr-rubric}:

\begin{equation}
\bar{\mathbf{s}}_{p,g}
=
\frac{1}{|\mathcal{M}|-1}
\sum_{e \in \mathcal{M}\setminus\{g\}}
\mathbf{s}_{p,g,e}
\label{eq:peer-aagr-rubric}
\end{equation}

where $\mathcal{M}$ denotes the set of evaluated agents. We similarly aggregate the overall quality score, evaluator confidence, and diagnostic counts across the three peer evaluations.

Finally, we compute both repository-level rankings and macro-level averages across all 15 benchmark repositories. Repository-level rankings identify the highest-performing agent for each individual project, whereas macro-level averages summarize overall performance across the benchmark. Because the repositories differ substantially in size, domain, and implementation characteristics, these aggregated results are intended to evaluate the robustness and consistency of each agent rather than establish superiority based solely on a single average score. This aggregation strategy enables \Name{} to identify agents that perform consistently across diverse repositories, exhibit high variability, or perform well only for specific project characteristics.

\section{Experimental Evaluation}
\label{sec:experimental-evaluation}

We evaluate \Name{} on four research questions:

\subsubsection*{\textbf{RQ1.}} Which agents generate broader test suites in terms of test-case volume and behavioral coverage across heterogeneous repositories?

\subsubsection*{\textbf{RQ2.}} Which agents act as more consistent and reliable evaluators in cross-agent peer assessment?

\subsubsection*{\textbf{RQ3.}} Which agents generate more runnable unit tests, and where do their generated suites break down across deeper adequacy dimensions?

\subsubsection*{\textbf{RQ4.}} To what extent does evaluator-reported confidence align with observed test quality, and can confidence serve as a useful proxy for adequacy?

\subsection{RQ1: Test-Suite Breadth and Behavioral Coverage}
\label{subsec:rq1}

We first examine the breadth of generated test suites under repository-only constraints. We define breadth from two perspectives: \emph{generation breadth}, measured by the number of generated test files and test cases, and \emph{behavioral breadth}, measured by the range of functional sectors covered across repositories. This distinction is important because an IDE may generate many individual test cases while still concentrating on a smaller set of modules, whereas another IDE may cover more files and functional areas with fewer cases.

\begin{table}[t]
\centering
\scriptsize
\setlength{\tabcolsep}{4pt}
\renewcommand{\arraystretch}{1.08}
\caption{Comparison of AI-generated unit test outputs across 15 projects. Each cell reports \#TestFiles/\#TestCases.}
\label{tab:testgen-comparison}
\begin{tabular}{p{0.34\linewidth}cccc}
\toprule
\textbf{Project} & \textbf{KIRO} & \textbf{ANTIGRAVITY} & \textbf{CURSOR} & \textbf{CS-4.5} \\
\midrule
Bus Tracker        & 12/151 & 1/10  & 7/113  & 4/96  \\
Cody                    & 14/74  & 6/52  & 2/26   & 3/153 \\
Dotfunding              & 4/48   & 2/7   & 5/65   & 4/48  \\
Du-Form-Automation      & 11/139 & 4/14  & 2/63   & 7/165 \\
Edushare                & 18/107 & 5/24  & 4/40   & 10/187 \\
Filr                    & 10/186 & 2/20  & 4/91   & 9/338 \\
Flashcard Generation    & 17/234 & 2/9   & 2/78   & 6/257 \\
Expense Tracker   & 12/288 & 5/75  & 4/89   & 6/282 \\
MediSched               & 10/145 & 8/155 & 5/194  & 5/160 \\
NUTRIMIND               & 13/247 & 3/56  & 6/92   & 4/138 \\
PatternCrafter          & 8/95   & 3/17  & 2/112  & 4/214 \\
Protiddhoni             & 21/382 & 3/29  & 4/99   & 4/99  \\
LIN                     & 5/40   & 1/14  & 3/41   & 1/239 \\
TiffinTime              & 13/119 & 9/157 & 4/53   & 6/158 \\
TravelHeaven            & 14/247 & 3/20  & 7/135  & 10/314 \\
\midrule
\textbf{Total}          & \textbf{182}/2502 & 57/659 & 61/1291 & 83/\textbf{2848} \\
\bottomrule
\end{tabular}
\end{table}

Table~\ref{tab:testgen-comparison} summarizes the generated unit test outputs for each repository, where each cell reports TestFiles/TestCases and the CS-4.5 column serves as the reference model against which the three IDEs---Kiro, Antigravity, and Cursor---are compared. Relative to this reference point of 83 files and 2848 cases, Kiro generates substantially more test files overall, with 182 files across 15 repositories, suggesting that this IDE achieves broader file-level and module-level coverage than the baseline. Antigravity and Cursor both generate fewer files than the baseline, with 57 and 61 files, respectively, and both also fall well short of the baseline on test-case volume, with 659 and 1291 cases against CS-4.5's 2848. This shows that among the three IDEs, only Kiro exceeds the baseline on file count, while none of the three IDEs approach the baseline on test-case volume, indicating that the CS-4.5 reference produces denser suites within fewer files than any of the IDEs achieve on their own terms. Thus, breadth depends on the measurement level relative to the baseline; Kiro alone surpasses CS-4.5 in file-level spread, whereas CS-4.5's case-level density remains unmatched across all three IDEs examined here.

\begin{figure}[t]
\centering
\includegraphics[width=\linewidth]{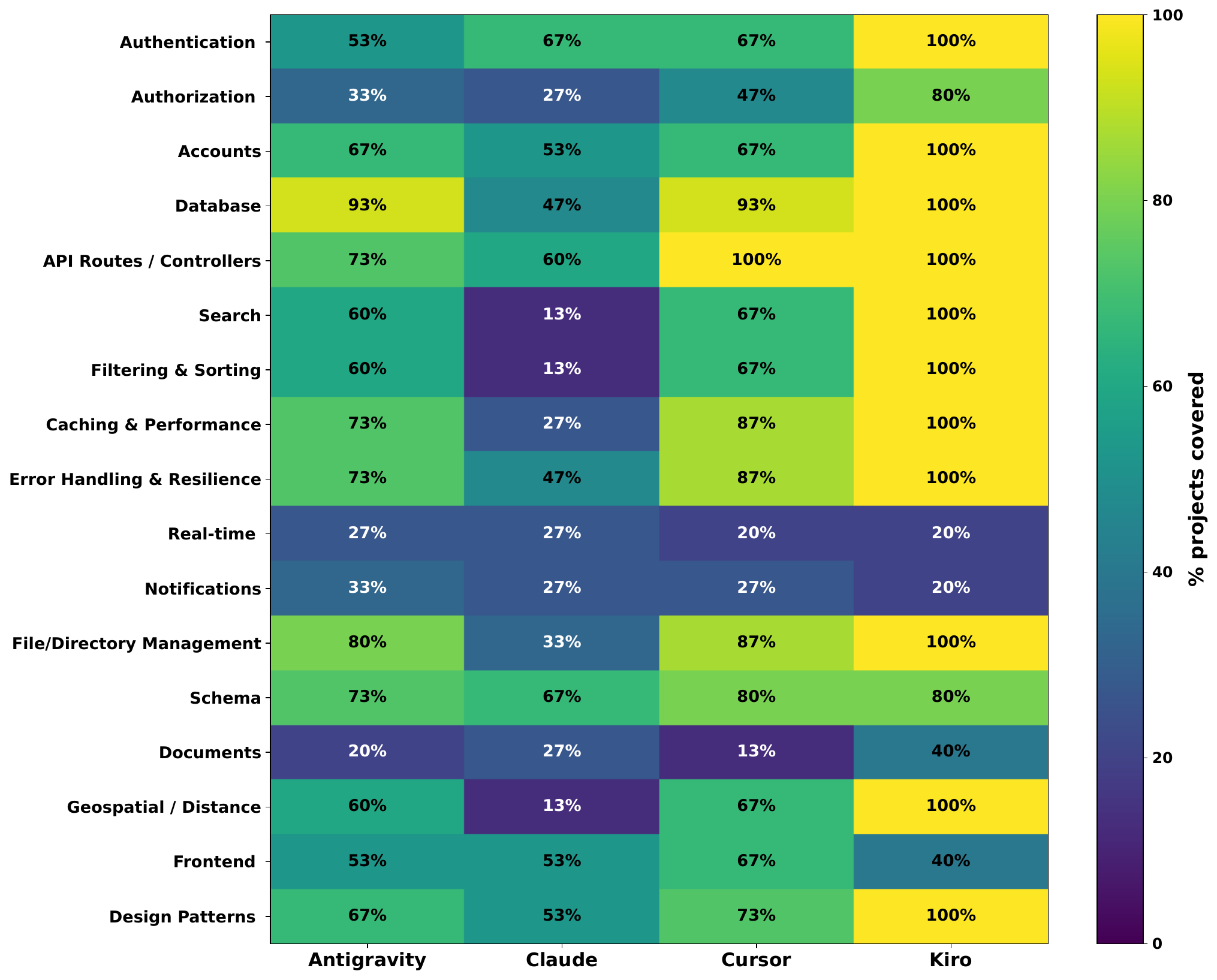}
\caption{Sector-level behavioral coverage breadth (\% of projects within each IDE's generated test suites that include at least one test targeting the given functional sector).}
\Description{A heatmap comparing the percentage of projects covered by Kiro, Antigravity, Cursor, and CS-4.5 across common and specialized functional sectors.}
\label{fig:sector-coverage}
\end{figure}

Figure~\ref{fig:sector-coverage} further compares behavioral breadth across functional sectors, again positioning CS-4.5 as the reference against which the three IDEs are read. Common backend- and validation-oriented behaviors, such as CRUD/database logic, API routes, authentication, validation, and error handling, are covered more frequently across the board, and the CS-4.5 baseline's coverage of these common sectors sits within this same well-covered band rather than standing apart from it. In contrast, specialized sectors such as real-time communication, notifications, document/PDF handling, and integration-heavy features receive weaker coverage generally, and the baseline shows the same pattern of thinner coverage in these harder sectors as the IDEs do. Kiro shows the broadest sector-level reach overall among the three IDEs, extending past the CS-4.5 reference in several specialized categories, while Cursor and Antigravity show more selective coverage patterns that fall below the baseline in most sectors. Read against the CS-4.5 reference point, this indicates that Kiro is the IDE most capable of matching or exceeding baseline behavioral breadth, while Cursor and Antigravity concentrate their coverage more narrowly than the baseline does.

\subsection{RQ2: Cross-Agent Evaluator Reliability}
\label{subsec:rq2}

Because Kiro, Antigravity, Cursor, and CS-4.5 all use the same underlying model, agreement between them cannot be interpreted as independent validation of correctness---shared-model behavior may itself drive agreement regardless of whether judgments are accurate. We therefore treat this analysis not as a measure of evaluator correctness, but as a probe of how much divergence in evaluation outcomes is attributable to each IDE's harness---its context handling, tool use, and workflow scaffolding---when the underlying model is held constant. In our leave-one-out protocol, each generated test suite is reviewed by the remaining agents, preventing self-evaluation and allowing us to compare evaluator agreement for Kiro, Antigravity, and Cursor relative to the CS-4.5 reference. Higher agreement indicates that an IDE's harness introduces little additional judgment divergence beyond the base model; lower agreement indicates that harness-specific scaffolding meaningfully shifts evaluation behavior away from the shared-model baseline.

\begin{figure}[t]
\centering
\includegraphics[width=\columnwidth]{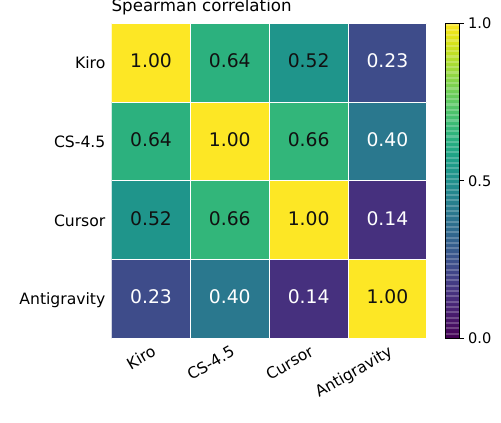}
\caption{Cross-evaluator agreement across peer reviewers relative to the CS-4.5 baseline, measured using Spearman $\rho$ over comparable generator-project instances.}
\Description{A heatmap of pairwise Spearman correlations among Kiro, Antigravity, Cursor, and CS-4.5 peer-evaluator scores.}
\label{fig:evaluator-agreement}
\end{figure}

Figure~\ref{fig:evaluator-agreement} shows that agreement is positive but uneven across the three IDEs, with CS-4.5 at the center of the strongest alignments in the matrix. Cursor shows the strongest agreement with CS-4.5, followed by Kiro, indicating that their harnesses diverge least from the base model's judgments. Antigravity shows the weakest agreement with CS-4.5, and this weakens further when paired with Cursor rather than CS-4.5---falling below either IDE's individual agreement with the baseline. Averaged across its pairings, CS-4.5 agrees more consistently with each IDE than the IDEs agree among themselves, supporting its use as a stable reference point despite the shared-model caveat. This suggests that harness design meaningfully shapes evaluation behavior even on identical underlying models, though the resulting agreement should be read as a measure of harness-induced consistency rather than evaluator correctness.

\subsection{RQ3: Runnability, Adequacy, and Failure Modes}
\label{subsec:rq3}

We next analyze which of the three IDEs generate more runnable and behaviorally adequate unit tests under repository-only constraints, reading their results against the CS-4.5 baseline throughout. Beyond output volume, we examine project-level overall scores, diagnostic failure signals, and rubric-level performance across five adequacy dimensions: runnability, assertion strength, logic and edge coverage, isolation and determinism, and maintainability.

\begin{table}[t]
\centering
\scriptsize
\setlength{\tabcolsep}{3pt}
\renewcommand{\arraystretch}{1.08}
\caption{Compact leaderboard across 15 projects (Overall score out of 100)}
\label{tab:leaderboard-compact-15}
\begin{tabularx}{\linewidth}{@{}Xccccc@{}}
\toprule
\textbf{Project} & \textbf{KIRO} & \textbf{ANTIGRAVITY} & \textbf{CURSOR} & \textbf{CS-4.5} & \textbf{Best} \\
\midrule
Flashcard Generation & 84.0 & \textbf{86.0} & 56.0 & 71.0 & ANTIGRAVITY \\
Du-Form-Automation & 38.0 & 52.0 & 54.0 & \textbf{75.0} & CS-4.5 \\
Expense Tracker & \textbf{98.0} & 56.0 & 74.2 & 59.0 & KIRO \\
PatternCrafter & \textbf{97.0} & 82.0 & 73.0 & 64.0 & KIRO \\
Dotfunding & \textbf{77.0} & 60.0 & 60.0 & 54.0 & KIRO \\
Protiddhoni & \textbf{98.0} & 71.0 & 64.2 & 83.0 & KIRO \\
LIN & 76.0 & 78.0 & 82.0 & \textbf{96.0} & CS-4.5 \\
Bus Tracker & \textbf{92.0} & 84.0 & 79.0 & 75.0 & KIRO \\
TiffinTime & 76.0 & \textbf{77.0} & 68.0 & 73.0 & ANTIGRAVITY \\
MediSched & 62.0 & 79.5 & 63.2 & \textbf{84.0} & CS-4.5 \\
Edushare & \textbf{84.0} & 36.0 & 81.8 & 70.0 & KIRO \\
Cody & \textbf{91.0} & 70.0 & 62.0 & 76.0 & KIRO \\
Filr & \textbf{82.0} & 75.0 & 70.0 & 80.0 & KIRO \\
TravelHeaven & \textbf{94.8} & 59.0 & 77.0 & 64.0 & KIRO \\
NUTRIMIND & \textbf{88.6} & 76.0 & 45.0 & 83.0 & KIRO \\
\midrule
\textbf{Win-rate (\#1)} & \textbf{10} & 2 & 0 & 3 & --\\
\bottomrule
\end{tabularx}
\end{table}

Table~\ref{tab:leaderboard-compact-15} presents the overall evaluation scores (out of 100) for each repository--IDE pair, where the Best column identifies the highest-scoring model for each repository and CS-4.5 again serves as the reference point against which Kiro, Antigravity, and Cursor are read. Kiro clearly outperforms the CS-4.5 baseline at the project level, ranking first on 10 of 15 repositories and beating CS-4.5's own three first-place finishes by a wide margin. Cursor never ranks first on any repository, placing it below both the baseline and the other two IDEs on this measure, while Antigravity ranks first on two repositories, still trailing the CS-4.5 baseline's three wins. Notably, CS-4.5's own three wins, on Du-Form-Automation, LIN, and MediSched, show that the baseline remains competitive on a meaningful subset of projects even though Kiro surpasses it overall. This indicates that Kiro is the IDE most capable of exceeding baseline project-level performance, while Cursor is the weakest of the three relative to CS-4.5 and Antigravity sits between them; repository structure and implementation style therefore shape how each IDE performs against the baseline.

\paragraph{Dominant failure modes.}

\begin{table}[t]
\centering
\scriptsize
\setlength{\tabcolsep}{3.5pt}
\renewcommand{\arraystretch}{1.08}
\caption{Failure mode rates by generator model across the 15 projects. Each cell reports \textit{\% of test files impacted} with \textit{events per 10 test files} in parentheses.}
\label{tab:failure-modes}
\begin{tabular}{@{}lcccc@{}}
\toprule
\textbf{Model} &
\textbf{Blocking} &
\textbf{Flaky} &
\textbf{Weak} &
\textbf{Missing} \\
\midrule
KIRO &
36.0\% (5.2) &
37.8\% (5.0) &
62.4\% (10.5) &
82.9\% (23.5) \\

ANTIGRAVITY &
75.5\% (11.7) &
61.9\% (12.7) &
92.3\% (25.4) &
99.4\% (72.2) \\

CURSOR &
45.1\% (7.7) &
48.0\% (8.1) &
93.1\% (20.4) &
98.8\% (38.1) \\

CS-4.5 &
52.3\% (12.2) &
58.9\% (12.2) &
81.8\% (22.9) &
99.2\% (61.4) \\
\bottomrule
\end{tabular}
\end{table}

Table~\ref{tab:failure-modes} summarizes the prevalence of four diagnostic failure categories---blocking errors, flakiness risks, weak assertions, and missing behavioral cases---for each of the three IDEs measured against the CS-4.5 baseline. Each cell reports the percentage of generated test files affected by a given failure mode, followed by the corresponding number of events per 10 test files in parentheses. Kiro shows a lower blocking-error rate and a lower flakiness rate than the CS-4.5 baseline, indicating that this IDE outperforms CS-4.5 on these two dimensions, while Antigravity and Cursor both show higher blocking-error rates than the baseline, with Antigravity showing the largest gap among the three IDEs. On weak assertions, Cursor and Antigravity both exceed the baseline rate, while Kiro is the only IDE with a notably lower weak-assertion rate than CS-4.5. Missing behavioral cases remain the most frequent failure signal for every IDE and for the baseline alike, with Kiro again showing the most favorable rate relative to CS-4.5 and the other two IDEs trailing behind it. This pattern suggests that Kiro's advantage over the baseline is broad across failure modes, whereas Antigravity and Cursor trail the baseline on most of these diagnostic signals even as all four remain executable or superficially plausible without fully encoding strong behavioral obligations.

\paragraph{Rubric level asymmetry.}

\begin{table}[t]
\centering
\scriptsize
\setlength{\tabcolsep}{3pt}
\renewcommand{\arraystretch}{1.08}
\caption{Rubric breakdown by generator model. Each cell shows mean rubric score with standard deviation in parentheses. $N$ is the number of test files generated across 15 projects.}
\label{tab:rubric-breakdown}
\begin{tabular}{@{}lcccccc@{}}
\toprule
\textbf{Model} & \textbf{$N$} & \textbf{R1} & \textbf{R2} & \textbf{R3} & \textbf{R4} & \textbf{R5} \\
\midrule
KIRO        & 182 & 4.41 (1.17) & 4.37 (1.24) & 4.52 (1.24) & 2.78 (2.37) & 0.83 (1.64) \\
ANTIGRAVITY & 57  & 3.80 (1.13) & 3.63 (1.26) & 3.82 (1.41) & 2.95 (1.99) & 1.37 (1.67) \\
CURSOR      & 61  & 3.45 (1.15) & 3.50 (1.23) & 3.66 (1.32) & 2.39 (1.88) & 0.95 (1.38) \\
CS-4.5      & 86  & 3.38 (0.98) & 3.36 (1.03) & 3.44 (1.25) & 2.64 (1.69) & 1.71 (1.55) \\
\bottomrule
\end{tabular}
\end{table}

Table~\ref{tab:rubric-breakdown} summarizes the mean scores for the five evaluation rubric dimensions across all generated test files, with standard deviations reported in parentheses, and the CS-4.5 row anchors the comparison across the three IDEs. On runnability and the two dimensions that follow it, R1 through R3, Kiro clearly exceeds the CS-4.5 baseline, Antigravity sits close to the baseline with a slight edge on R1 through R3, and Cursor sits slightly above the baseline on these same three dimensions. The pattern reverses on the later rubric dimensions; on R4, both Kiro and Antigravity exceed the baseline while Cursor again trails it, and on R5, the CS-4.5 baseline achieves the highest score of all four models, ahead of Kiro, Antigravity, and Cursor alike. This indicates that although Kiro is the strongest of the three IDEs on the earlier structural rubric dimensions, none of the three IDEs match the CS-4.5 baseline on maintainability, which remains the one dimension where the reference model leads the full comparison.

\paragraph{Qualitative failure example.}
A representative failure appears in \texttt{TravelHeaven}, where a negative path service test used a \texttt{try/catch} structure.

\begin{lstlisting}[language=JavaScript]
test('should throw for missing item', async () => {
  try {
    await approvalService.approveItem('missing-id');
  } catch (e) {
    expect(e.code).toBe(404);
  }
});
\end{lstlisting}

Although executable, this test can pass when \texttt{approveItem} does not throw, because no assertion is executed. Evaluators therefore marked it as both a weak assertion and a missing behavioral obligation. This example illustrates that runnability alone does not guarantee behavioral adequacy, a point worth keeping in mind when reading any IDE's rubric scores against the CS-4.5 baseline in the tables above.

\subsection{RQ4: Confidence-Quality Alignment}
\label{subsec:rq4}

\begin{figure}[t]
\centering
\includegraphics[width=\linewidth]{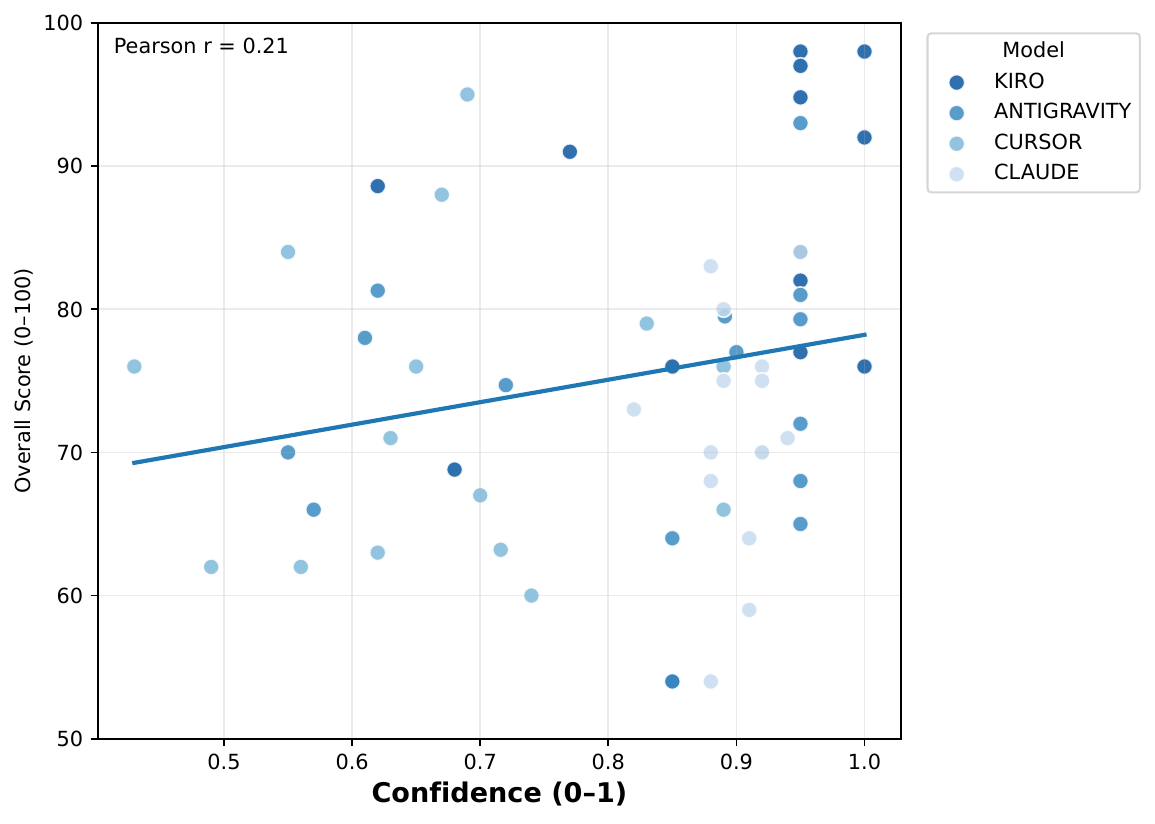}
\caption{Confidence vs. Overall score (Pearson $r = 0.21$).}
\Description{A scatter plot of evaluator confidence against overall quality score for Kiro, Antigravity, Cursor, and CS-4.5, showing a weak positive relationship.}
\label{fig:confidence-vs-overall}
\end{figure}

We finally analyze whether evaluator-reported confidence aligns with observed test quality, and whether this relationship holds consistently for the three IDEs when read against the CS-4.5 baseline. Figure~\ref{fig:confidence-vs-overall} shows a weak positive association between evaluator confidence and overall test quality across all four models pooled together (Pearson $r = 0.21$, 95\% CI $[-0.05, 0.44]$, $p = 0.11$, $n=60$). The CS-4.5 points cluster in the middle-to-lower confidence range with overall scores that are broadly comparable to the other three IDEs at similar confidence levels, rather than standing out as systematically higher or lower. Kiro and Antigravity points tend to sit at the higher end of the confidence axis while also reaching some of the highest overall scores in the plot, whereas Cursor and the CS-4.5 baseline show more dispersion at the same confidence levels, including several points where confidence is high but the overall score remains closer to the middle of the range. This indicates that high confidence does not reliably distinguish the IDEs from the CS-4.5 baseline, since points from all four models overlap substantially across the confidence axis. Confidence therefore provides only a limited proxy for adequacy relative to the baseline; while higher-confidence judgments tend to be somewhat better on average across Kiro, Antigravity, Cursor, and CS-4.5 alike, substantial score dispersion remains even among predictions with similar confidence values. In practice, confidence may still be useful for coarse prioritization or triage when comparing an IDE's output against the CS-4.5 baseline, but it is not reliable enough to replace structured rubric-based assessment for that comparison.

\section{Limitations}

This study is an initial investigation into how modern AI-assisted IDEs perform at unit test generation, and several limitations point toward directions for future work.

First, each IDE was tested with a single underlying LLM rather than multiple LLMs, so our results reflect specific IDE-model pairings rather than the full range of possible configurations. Testing multiple LLMs per IDE would help separate the IDE's own contribution from the model's.

Second, our evaluation relies entirely on cross-agent LLM peer assessment, without human feedback. The leave-one-out protocol, multiple independent reviews, and checkable diagnostic signals reduce calibration and self-bias effects, but no human-annotated validation of rubric scores was performed.

Third, the benchmark uses 15 student-developed repositories rather than production-ready projects. These are smaller and less complex than industrial codebases, so it remains unclear whether the observed failure patterns hold at real-world scale and complexity.

Fourth, all IDEs were tested under a single zero-shot, repository-only prompting paradigm. We did not explore few-shot examples, iterative refinement, or external documentation. Since IDE performance may vary with prompting strategy, testing multiple paradigms would clarify whether the weaknesses observed here are intrinsic to current IDEs or artifacts of our setup.

Taken together, this study shows that current LLM-based IDEs do not yet produce reliably adequate unit tests under repository-only, zero-shot conditions.

\section{Conclusion}

This paper presented \Name, a repository-grounded framework for evaluating how AI-assisted IDEs perform at unit test generation under strict repository-only constraints, using Claude as the baseline against which Kiro, Antigravity, and Cursor were compared. Across 15 heterogeneous repositories, we find a clear execution-adequacy gap that holds for both the baseline and the IDEs: tests often run successfully but still contain weak assertions, missing edge cases, isolation issues, and maintainability problems. Kiro shows the strongest overall project-level performance, exceeding the Claude baseline on most repositories, while Claude generates denser test suites than any of the three IDEs. Overall, our results show that AI-generated tests, whether from the baseline or the IDEs built around it, should be evaluated beyond pass rates and coverage, using structured adequacy criteria and diagnostic failure analysis, and the limitations above point to concrete ways this evaluation can be broadened in future work.

\balance
\bibliographystyle{ACM-Reference-Format} 
\bibliography{reference}

\end{document}